# DC-assisted microwave quenching of $YBa_2Cu_3O_{7-\delta}$ coplanar waveguide to a highly dissipative state

N.T.Cherpak[1], A.A.Lavrinovich[1], A.I.Gubin[1] and S.A.Vitusevich[2, a]

[1] O. Usikov Institute for Radiophysics and Electronics NAS of Ukraine, 12 ac. Proskura str., 61085 Kharkiv, Ukraine

[2] Peter Grünberg Institute, Forschungszentrum Juelich, 52425 Juelich, Germany

Abstract

The paper reports on finding the effect of a strong change in the microwave losses in an HTS-based coplanar waveguide (CPW) at certain values of the input power $P_{in}$ and direct current $I_{dc}$. CPW on the basis of 150 nm thick $YBa_2Cu_3O_{7-\delta}$ epitaxial film on a single crystal MgO substrate was studied experimentally. A sharp and reversible transition of the CPW into a strongly dissipative state at the certain meanings of $P_{in}$ and $I_{dc}$ depending on temperature was observed. Apparently the effect can be explained by self-heating of HTS structure caused by magnetic flux flow under the joint influence of MW and DC.

Since discovery of high-temperature superconductors (HTS) in 1986 [1] a lot of attention has been given to studying nonlinear properties of HTS in microwave (MW) fields [2-4]. Peculiarities of the MW nonlinear response of the superconductors are due to varieties of the interaction mechanisms of electromagnetic fields with these materials, and are equally important for both Physics and applications of HTS.

[a] Electronic mail: s.vitusevich@fz-juelich.de

Important characteristic of a superconductor in the MW field is the surface impedance $Z_s = E_s / H_s$, where $E_s$ and $H_s$ are the electric and magnetic fields on the surface of the superconductor. In the case of local electrodynamics the impedance $Z_s$ can be expressed in terms of the complex conductivity σ as $Z_s = (i\omega\mu_0 / \sigma)^{1/2} = R_s + iX_s$, where ω is the MW field angular frequency, $\mu_0 = 4\pi * 10^{-7}$ H / m is the permeability of free space [2].

Increase of the field amplitude $H_\omega$ (or power $P_\omega$) can change the conductivity σ, which leads to a dependence of $Z_s$ on $H_\omega$. There are a number of nonlinearity mechanisms responsible for $Z_s(H_\omega)$. However, for high-quality HTS films, which are superconductors of the second kind, the most pronounced, apparently, is the mechanism associated with the formation of vortices and flux flow [2-9].

In contrast to the infinite conductivity of superconductors for direct current (DC) $I_{dc}$ smaller than the critical current $I_c$, MW fields produce heat in a superconductor at any value of the MW current $I_\omega$. Effect of heating increases substantially when the magnetic flux flow resistance kicks in. These effects lead to an increase in MW loss at $P_\omega$ values typical for practical devices [2]. The temperature of the superconductor can be noticeably increased as compared with a temperature provided by cryogenic system. Microwave heating may affect the whole or selected areas of the superconductor, and thus may be a global or local heating.

Soon after the discovery of high-temperature superconductivity [1] an idea of creating MW power limiter using dependence $R_s (P_\omega)$ was suggested and some of the first experiments were carried out with a planar transmission line [10, 11]. This approach was originally developed in [12,13]. Despite that the possibility of constructing such a limiter was confirmed by [12, 13], serious obstacles were found in the implementation of practical devices: the difficulty to control the input power handling $P_{\omega(in)}$ [12] and the lack of stability of planar transmission line to the breakdown upon the transition to a strongly dissipative condition [13]. In [14] it was proposed to control the nonlinear impedance of a coplanar waveguide created on the basis of HTS films by using DC bias current. This possibility in principle was demonstrated, however irreversible changes in the properties of HTS film in the transmission line remained an unresolved problem.

In this paper we report on discovery of the effect of a strong and abrupt change in the microwave losses in a section of an HTS-based coplanar line at certain values of the input MW power $P_{\omega(in)} = P_{in}$ and direct current $I_{dc}$. Here, no irreversible changes in HTS film were observed.

For the experimental study a coplanar waveguide (CPW) on the basis of 150 nm thick $YBa_2Cu_3O_{7-\delta}$ epitaxial film on a single crystal MgO substrate was fabricated by photolithography. The HTS film produced by THEVA (Germany) had the following characteristics: $T_c = 86.5$ K, $J_c = 3.6 \cdot 10^6$ A/cm$^2$ at $T = 77$ K. CPW cross-section dimensions were $a = 0.186$ mm, $w = 0.1$ mm, and $h = 0.5$ mm (Fig. 1a).

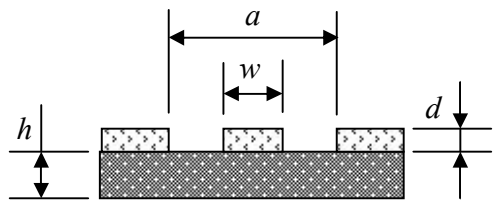

Fig.1a. Cross-section of coplanar waveguide.

CPW is a 16.81 mm long straight section with gold-plated pads, through which the feed-in and readout of the microwave signal and, simultaneously, DC transmission were implemented by us-ing the integrated planar bias tees (IPBT). A magnetron source operated at $f = 9.24$ GHz, with the output power adjustable in the range of 0 to 13 watts. Continuous signal from the magnetron was modulated by the square-wave pulse generator with the following characteristics: pulse duration $\tau_i = 5$ μs, the pulse repetition period $T = 40$ μs (repetition frequency $f_r = 2.6 * 10^4$ Hz). CPW was placed into a chamber filled with helium gas. The chamber was cooled by liquid nitrogen. The block diagram of the experimental set-up with the DC source (DCS) is shown in Fig. 1b.

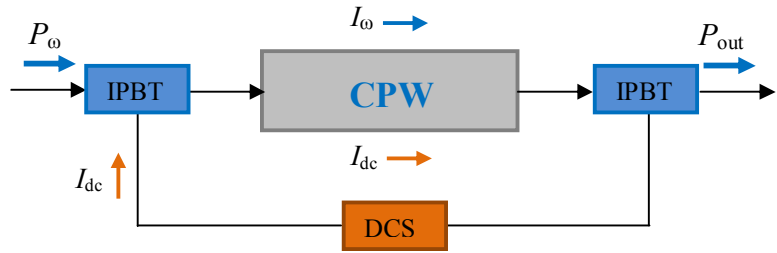

Fig.1b. Block diagram of the experimental set-up: IPBT is an integrated planar bias tee, CPW is a coplanar waveguide and DCS is DC source.

In the experiments the insertion loss of CPW, $IL = 10 \lg (P_{out} / P_{in})$, was measured versus both temperature of the cooled chamber and the bias DC current $I_{dc}$ when a fixed level of the input MW signal $P_{in}$ was applied. Here $P_{out}$ is the power measured at the output of CPW.

Figure 2 shows dependence of the insertion loss $IL$ on temperature $T$ for several values of $P_{in}$. We see that at a small $P_{in}$ the function $IL(T)$ is monotonic. However, with increasing $P_{in}$, along with a general increase in CPL losses, a peculiar feature in the form of steps appears in the dependence $IL(T)$. The overall increase in losses is caused by an increase of the surface resistance $R_s(H_\omega)$ or $R_s(P_\omega)$ [2] ($H_\omega$ and $P_\omega$ are the MW magnetic field amplitude and the power in the CPW, correspondingly). The appearance of steps, possibly, is due to the formation of thermal domains [15, 16] and/or the manifestation of the wave reflection from the input and output of the CPW [13].

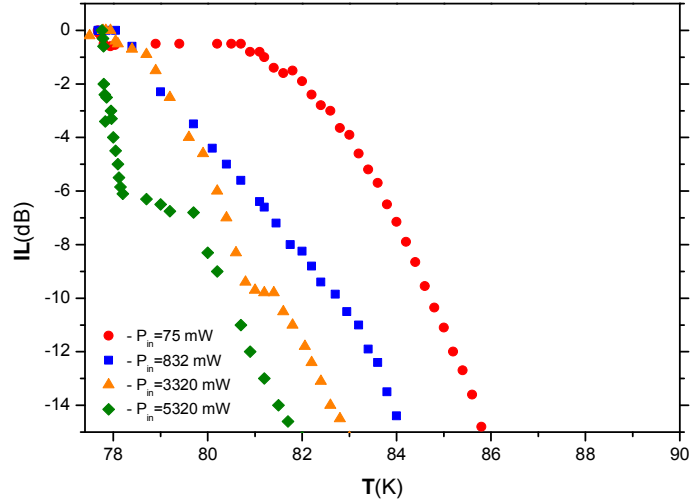

Fig.2. Dependence of the insertion loss in CPW on a temperature at different input powers of 5 μs long MW pulses.

This dependence noticeably changes when different currents $I_{dc}$ were passed through the CPW (Fig. 3). It's seen that if the values of $P_{in}$ and $I_{dc}$ are fixed, there is a sharp transition of the CPW into a strongly dissipative state at a certain temperature with the jump in the losses $IL$ of almost three orders of magnitude (in this particular case for = 40 mA). Avalanche growth of the microwave loss $IL$, i.e. microwave quenching stimulated by DC, remained on the $IL$ $(P_{in})$ curves at $I_{dc}$ = const (not shown in the figure) and the $IL$ $(I_{dc})$ curves at $T$ = const and different values of $P_{in}$ (Fig. 4). The quenching phenomenon takes place at a specific values of $I_{dc} = I^*$. Fig. 4 shows that with increasing $P_{in}$, $I^*$ values decrease. It follows that for realization of the quenching process both current components, microwave $I_\omega$ and DC $I_{dc}$ currents, play a role. It makes sense to find out whether the contributions of these components are additive in the observed effect of quenching.

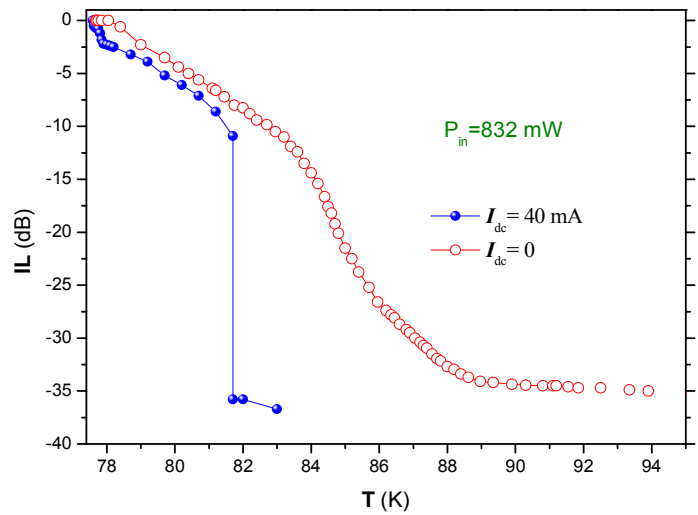

Fig.3. Dependence of the insertion loss in CPW on temperature at 5μs-long pulse of $P_{in}$ = 832 mW without ($I_{dc}$ = 0) and with ($I_{dc}$ = 40 mA) DC.

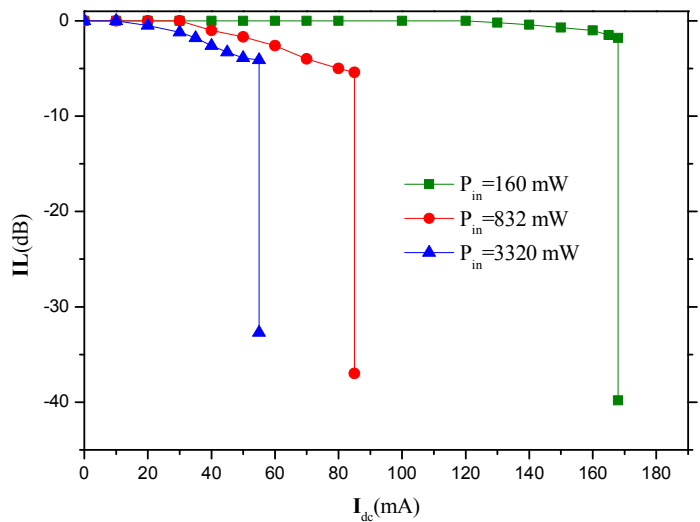

Рис.4. Dependence of the insertion loss in CPW on direct current at $T$ = 77 K at different values of the input power $P_{in}$ modulated by 5 μs-long pulses.

In [17] it is shown that in $YBa_2Cu_3O_{7-\delta}$ thin film microbridges, similar to the HTS planar structure studied in the present paper, a switching effect (quenching) is observed in the dissipative state when CPW is exposed to millisecond-long high DC pulses. The authors have measured the current-voltage curves in the microbridges with widths lower than the thermal diffusion length. Such an approach allowed finding that the flux-flow resistivity is microbridge-width independent, when strong width dependence of the quenching current density is observed. The authors [17] concluded that for high current densities varying in the millisecond range, the intrinsic quenching mechanism is self heating caused by conventional flux-flow effects. However the authors do not exclude the relevance of other quenching mechanisms for other conditions.

If we assume that the quenching mechanism in MW field remains the same, it makes sense to evaluate the ratio of DC and MW current densities, $J^*_{dc}$ and $J_\omega$, triggering the quenching effect, as a function of the input power $P_{in}$. Although the central conductor of CPW is not a superconducting channel where $J_{dc}$ would be a function of the position within the cross-section of the strip, it can be assumed that in our case that $J^*_{dc}$ varies weakly over the strip cross section $S$, as film is already in the resistive or pre-resistive state under the action of MW field alone. Therefore $J^*_{dc} \approx I_{dc}/S$. It can be noted that at low $P_{in}$ the characteristic value of $J^*_{dc}$ decreases with increased $P_{in}$ more sharply than at higher levels of $P_{in}$ (Fig. 5). This feature might be explained by inhomogeneous distribution of the MW current in the cross-section of the CPW and the partially additive nature of $J_{dc}$ and $J_\omega$ effects on the HTS film state. The role of $J_\omega$ non-uniformity in CPW, as seen in Fig.5, decreases with increasing $P_{in}$. The maximum value of the sum of the current densities $J_{dc}$ and $J_\omega$ always remains below the value of the passport density of the critical current $J_c = 3.6 \cdot 10^6$ A/cm$^2$. For homogeneous MW current distribution we could expect that the sum $J^*_{dc} + J_\omega$ is independent on $P_{in}$ if the nature of both $J^*_{dc}$ and $J_\omega$ effects on the CPW transmission is the same.

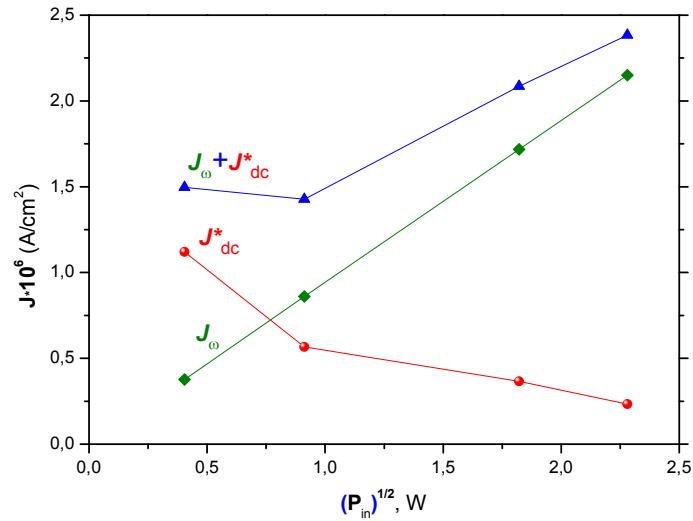

Fig.5. Current densities $J^*_{dc}$, $J_\omega$ and $J^*_{dc}+J_\omega$ as functions of the square root of the input power $P_{in}$ modulated by 5 μs-long pulses.

In summary, we have experimentally found that at certain values of the input power $P_{in}$ in the HTS coplanar waveguide an abrupt DC-assisted microwave quenching switches the HTS film into a strongly dissipative state. We speculate that the observed effect can be explained by self-heating of HTS structure caused by magnetic flux flow under the joint influence of MW and direct currents.

**Acknowledgement**. We are grateful to Dr. A.V.Velichko from the University of Nottingham, UK for helpful discussion and useful comments.